\documentclass[a4paper,conference]{IEEEtran}
\IEEEoverridecommandlockouts
\usepackage{cite}
\usepackage{amsmath,amssymb,amsfonts}
\usepackage{algorithmic}
\usepackage{graphicx}
\usepackage{textcomp}
\usepackage{xcolor}
\usepackage{multirow}
\usepackage{makecell}
\def\BibTeX{{\rm B\kern-.05em{\sc i\kern-.025em b}\kern-.08em
    T\kern-.1667em\lower.7ex\hbox{E}\kern-.125emX}}
    
\begin{document}

\title{Parkinson's Disease Detection with Ensemble Architectures based on ILSVRC Models
}

\author{\IEEEauthorblockN{Tahjid Ashfaque Mostafa}
\IEEEauthorblockA{\textit{Department of Computing Science} \\
\textit{University of Alberta}\\
Edmonton, Canada \\
tahjid@ualberta.ca}
\and
\IEEEauthorblockN{Irene Cheng}
\IEEEauthorblockA{\textit{Department of Computing Science} \\
\textit{University of Alberta}\\
Edmonton, Canada \\
locheng@ualberta.ca}
\and
}

\maketitle
\begin{abstract}
In this work, we explore various neural network architectures using Magnetic Resonance (MR) T1 images of the brain to identify Parkinson's Disease (PD), which is one of the most common neurodegenerative and movement disorders. We propose three ensemble architectures combining some winning Convolutional Neural Network models of ImageNet Large Scale Visual Recognition Challenge (ILSVRC). All of our proposed architectures outperform existing approaches to detect PD from MR images, achieving upto 95\% detection accuracy. We also find that when we construct our ensemble architecture using models pretrained on the ImageNet dataset unrelated to PD, the detection performance is significantly better compared to models without any prior training. Our finding suggests a promising direction when no or insufficient training data is available.
\end{abstract}


\section{Introduction}
In recent years, Computer aided diagnosis systems based on brain imaging have shown merits in the diagnosis of Parkinson's Disease(PD), with the objective to detect PD by automatic recognition of patterns that characterize it. PD is the second most common neurodegenerative disorder and the most common movement disorder affecting the elderly next to the Alzheimer's disease~\cite{mhyreboyd}. The root cause of PD is thought to be the loss of nerve cells (neurons) in Substantia Nigra part of Basal Ganglia, which is one of the major regions of the human brain located deep within the cerebrum, below cerebral cortex. Dopamine, the neurotransmitter of the brain, is produced in this region, which facilitates the communication between neurons. This communication is essential for coordination of body movement and a shortage of dopamine hampers it; leading to PD. 

PD has been associated with neurological symptoms like speech impediments, olfactory dysfunctions, sleep disorders, autonomic dysfunctions, fatigue and  balance issues like tremors, Bradykinesia, postural instability, rigidity of the limbs, impaired gait etc. It has been clinically studied and defined for a long time, but the exact mechanisms leading to PD are still not properly identified~\cite{unclear_Cause}. In majority of the cases, PD is diagnosed with the manifestation of motor symptoms. But these symptoms might not become apparent until 50 to 70\% of neurons have been damaged~\cite{5070neurons}, which is too late for any sort of preventive measures. Even though a guaranteed cure for PD has not been discovered yet, early detection might offer an opportunity for slowing or stopping the progression of the disease. There are also new forms of treatment like Exenatide~\cite{exenatide}, which show promising results with cases where PD was detected in the initial stages. 

One of the techniques that has been found to be successful in detecting neurodegenerative diseases with cognitive impairments~\cite{medicalimaging1}~\cite{medicalimaging2} is the analysis of the structural changes in the brain using Medical Imaging techniques. Specifically, Magnetic Resonance (MR) images, which posses high contrast and resolution within soft tissue have been found to provide better performance in brain structure analysis. 

In this work, we propose three ensemble architectures to identify PD from MR images of the brain and analyze whether models pre-trained on unrelated Imagenet \cite{imagenet} dataset perform better than models without any prior training in detecting PD. We achieve over 90\% detection accuracy for all three of our architectures with the highest being 95.15\%, which is better than existing models using similar data.  We also found that using models pretrained on the Imagenet dataset to construct our ensemble architecture yields much better performance than using untrained models, even though the training data is unrelated to PD.
\section{Background and Related Works}
A multitude of Machine Learning (ML)~\cite{unclear_Cause,relatedwork1,paz,relatedwork2} and Deep Learning (DL)~\cite{choi,relatedwork3} based approaches have been introduced for the detection of Parkinson's Disease. Focke et al.~\cite{relatedwork1} extracted Gray Matter (GM) and White Matter (WM) from MR images and fed them to a SVM Classifier for PD detection achieving 39.53\% and 41.86\% classification accuracy for GM and WM respectively. Radial Basis Function Neural Network (RBFNN) was used by Pazhanirajan et al.~\cite{paz} for PD classification. 
Babu et al.~\cite{babu} achieved a 87.21\% accuracy in classifying PD using GM  with a Computer Aided Diagnosis (CAD) system. They identified Superior Temporal Gyrus as a potential biomarker which plays a vital role for PD. 

Choi et al.~\cite{choi} achieved an accuracy of 96\% using SPECT imaging with Convolutional Neural Network [CNN]. Although their accuracy was very high, SPECT Imaging is invasive and not very popular as it requires injecting a radioactive tracer into the patient. Around 100 times more MRI scans were performed compared to SPECT over one year period in the NHS operation in England. Thus the SPECT approach seems to be impractical for normal medical use due to limited sample size, despite its reported high accuracy. The dataset is also class imbalanced since about 69\% of the data is from PD patients. Class imbalance causes the models to over classify the majority class~\cite{classimbalance}.

To detect PD from resting-state functional MRI (rsf-MRI), which detects subtle changes in blood oxygenation level, whereas Structural MRI (sMRI) only captures the anatomical details and ignores all activity, Long et al.~\cite{long} used a ML based approach and they achieved 87\% classification accuracy, but the dataset used by them was very small. Rana et al.~\cite{rana} used a SVM for classification with t-test feature selection on WM, GM and Cerebrospinal Fluid (CSF) achieving 86.67\% accuracy for GM and WM and 83.33\% accuracy for CSF. In another work~\cite{rana1}, the authors used the relation between tissues instead of considering the tissues separately and achieved an accuracy of 89.67\%. 

Among the various regions in the brain, the Substantia Nigra (SN) region has significant correlation with PD according to Braak’s neuroanatomical model of Parkinson’s Disease~\cite{braak} and it is often used as a Region Of Interest (ROI) in PD identification. But we will not be using this region of the brain for our analysis, instead we consider GM and WM only, since we found that the detection accuracy increased drastically using GM and WM in our previous work~\cite{firstpaper}. 

\begin{figure}[t]
    \centering
    \begin{tabular}{c}
        \includegraphics[width=0.45\textwidth]{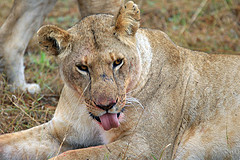}\\
        \small{ (a) Animal-Beast-Chordate-Vertebrate-Mammal-Placental-}\\\small{Carnivore-Feline-Big Cat-Lion} \\
        \includegraphics[width=0.45\textwidth]{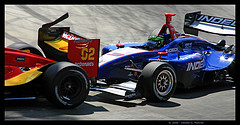}\\
        \small{ (b) Artifact-Instrumentation-Container-Wheeled Vehicle-}\\\small{Self Propelled Vehicle-Motor Vehicle-Car/Automobile-Race Car}     
    \end{tabular}
\caption{Sample Images from ImageNet~\cite{imagenet} dataset and their position in the WordNet~\cite{wordnet,wordnet1} Hierarchy}\label{imagenetimages}

\end{figure}

\begin{figure}[htbp]
\centering
\includegraphics[width=0.5\textwidth]{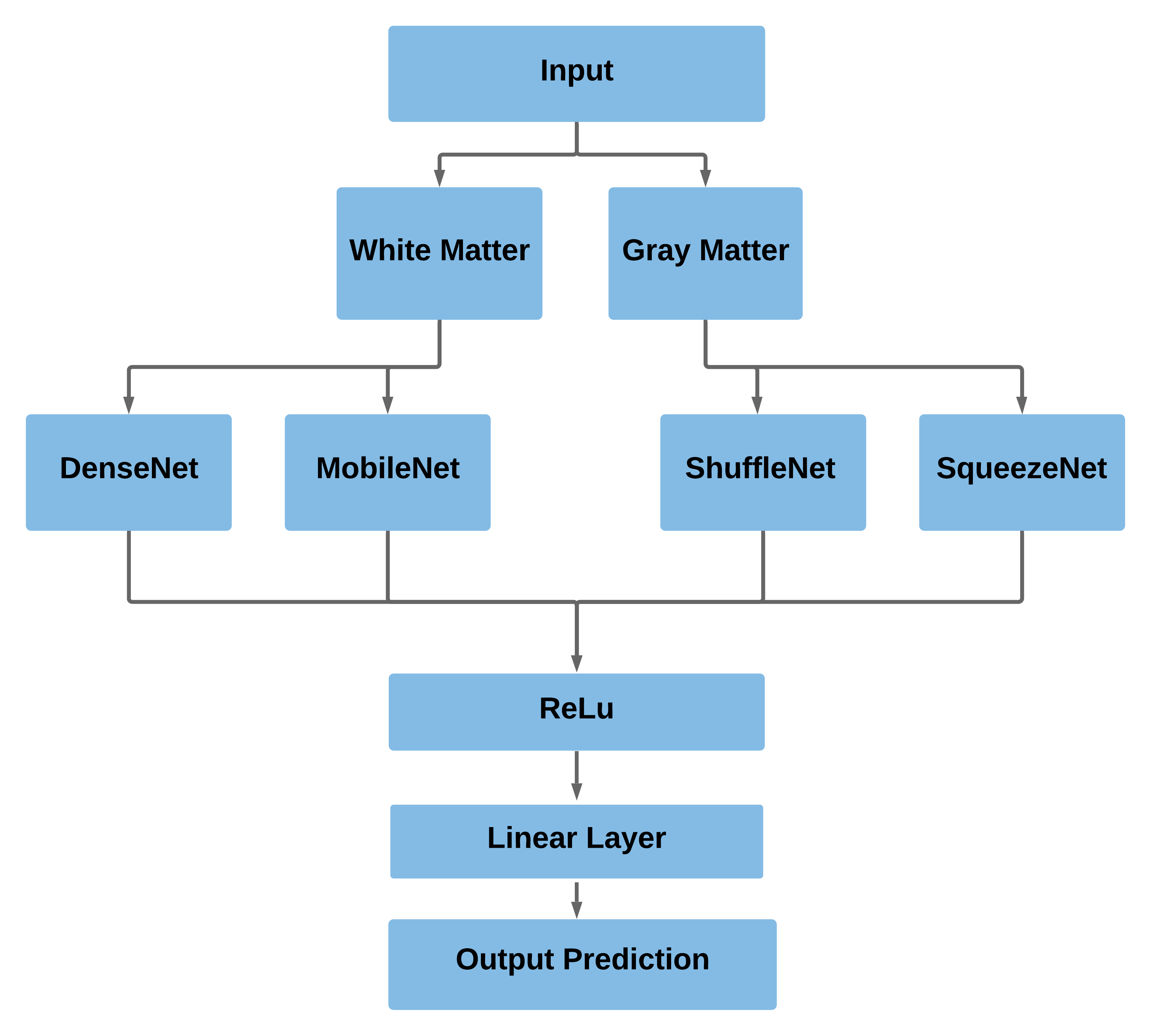}
\caption{Ensemble architecture from previous experiment~\cite{firstpaper}} \label{oldarc}
\end{figure}

ImageNet~\cite{imagenet} is one of the well known image datasets for computer vision. It is unrelated to PD detection. ImageNet is organized according to the WordNet~\cite{wordnet,wordnet1} hierarchy. WordNet is one of the largest lexical database of English words with nouns, verb, adjectives etc organised into "synonym sets" or "synsets", which are sets of cognitive synonyms. A "synset" describes a meaningful concept with multiple words or word phrases. WordNet contains more than 100,000 synsets, with more than 80,000 being nouns. Currently ImageNet labels images using only the nouns from WordNet. Each node of WordNet hierarchy is represented by thousand image samples in ImageNet on average. The ImageNet Large Scale Visual Recognition Challenge(ILSVRC)~\cite{imagenetcontest} evaluates the performance of various algorithms for object detection and image classification on the ImageNet dataset. The challenge has 1000 object categories, with the categories containing both internal and leaf nodes of ImageNet, but they do not overlap.
Fig.~\ref{imagenetimages} shows two sample images from the ImageNet dataset and their positions in the WordNet hierarchy.
Kornblith et al.~\cite{transfer} proposed that models performing well on the ILSVRC also perform better when they are applied on other datasets. 
In our previous work~\cite{firstpaper}, we tried a combination of multiple neural networks trained on PD data to create ensemble architectures for detecting PD, as shown in Fig. ~\ref{oldarc}. In this work, we built ensemble architectures, composed of ILSVRC models pretrained using non-PD related ImageNet images. We then used MR images to validate the effectiveness of these architectures. We seperated WM and GM from MR scans of the brain and passed them through our architectures. Experimental results showed that when using GM and WM, ensemble architectures with models pretrained on ImageNet data achieve better performance in detecting PD. 

The objectives of this paper are two-folds, 
(1) We want to know whether Kornblith's hypotheses and our observation from previous work holds true if we transfer those models learnt on completely unrelated data to detect PD. 
(2) The detection accuracy using sMRI was lacking in current methods and none of the methods we looked into were using ensemble architectures for PD detection. We want to demonstrate that creating ensemble architectures combining different characteristics of deep learning models has potential to give us better results. Our approach is described in the next section.

\section{Proposed Method}
\subsection{Data}
We used Parkinson Progression Markers Initiative (PPMI) dataset~\cite{ppmi} for our experiments, which consists of T1-weighted sMRI scans for 568 PD and Healthy Control(HC) subjects. We only chose 445 subjects and discarded the rest due to structural anomalies during preprocessing steps. There was a class imbalance in the resulting data with 299 PD and 146 HC subjects. To balance the data, we collected 153 HC T1-weighted sMRI scans from the publicly available IXI dataset~\cite{ixi}. The final dataset was class balanced with 598 subjects. The demographic for the dataset is presented in Table~\ref{demotable}. 
\begin{table}[htbp]
\centering
\setlength{\tabcolsep}{5pt}
\renewcommand{\arraystretch}{1}
\caption{Demographic Data}\label{demotable}
\begin{tabular}{l|c|c|c}
 &  PD & HC & Average\\
\hline
Age(Years) &  $62.0 \pm 9.54$ & $49.2 \pm 16.9$ & $55.6 \pm 15.1$\\\hline
Sex (Male / Female) &  189 / 110 & 172 / 127 & 361 / 237\\
\end{tabular}
\end{table}

\subsection{Preprocessing}
The scans we had come from different machines, so there were dimensional and morphological disparities between the data. We had to standardize the data to a common format to make it comparable. All scans were resized to the same dimensions. For preprocessing, Statistical Parameter Mapping (SPM12)~\cite{spm,spm1} and Computational AnatomTtoolbox (CAT12)~\cite{cat12} were used.

\begin{figure}[htbp]
\centering
\includegraphics[width=0.5\textwidth]{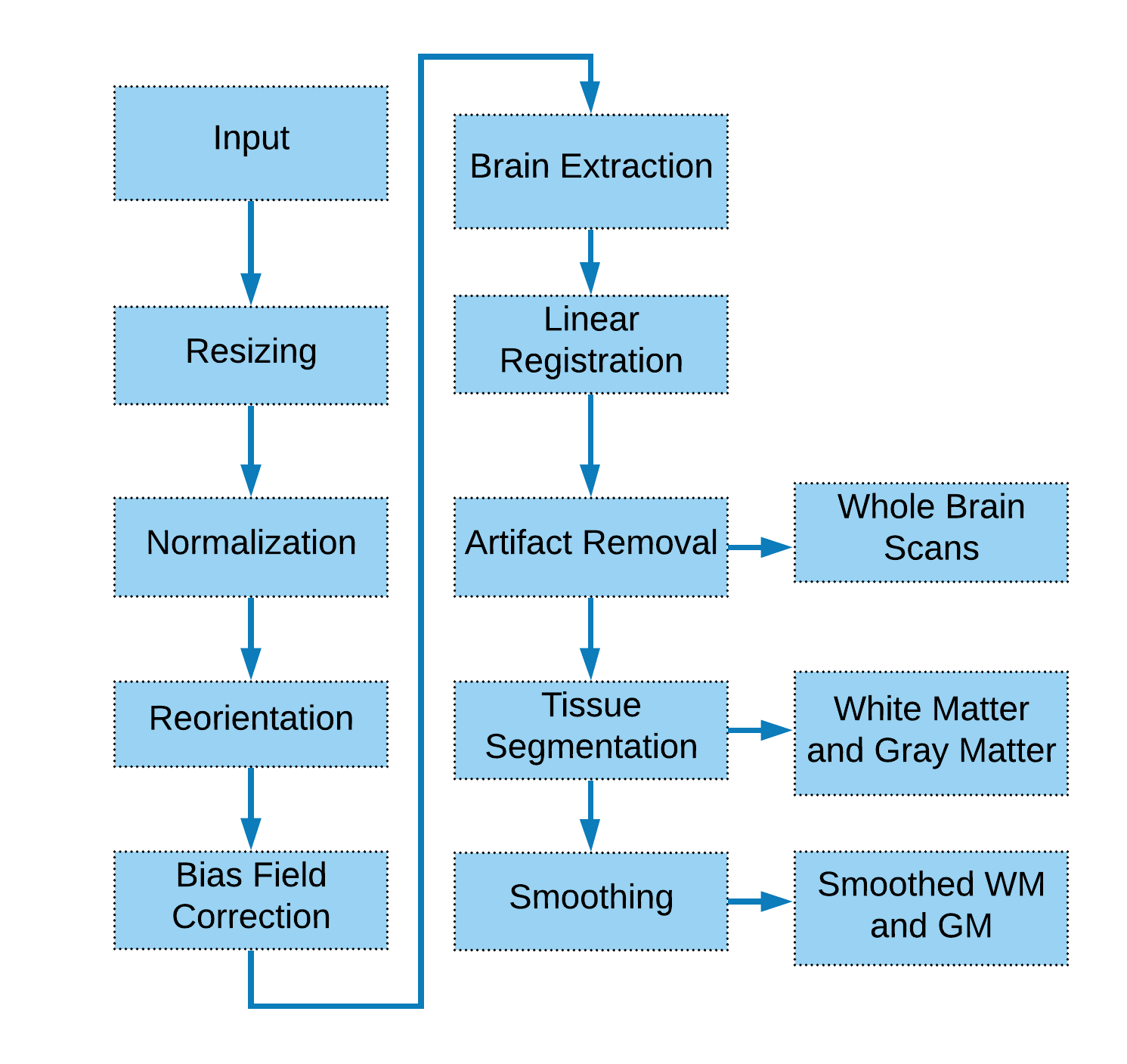}
\caption{Preprocessing Pipeline} \label{fig1}
\end{figure}

\begin{figure}[!t]
    \centering
    \begin{tabular}{r}
        \includegraphics[width=0.45\textwidth]{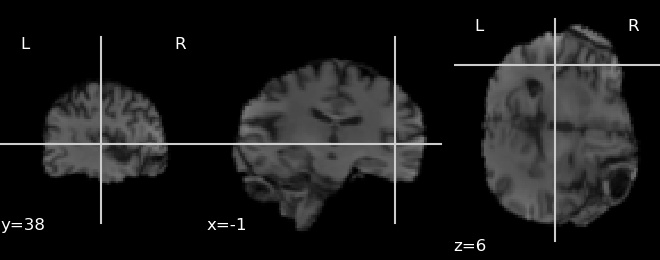}\\
        \multicolumn{1}{c}{\small (a) Whole Brain} \\
        \includegraphics[width=0.45\textwidth]{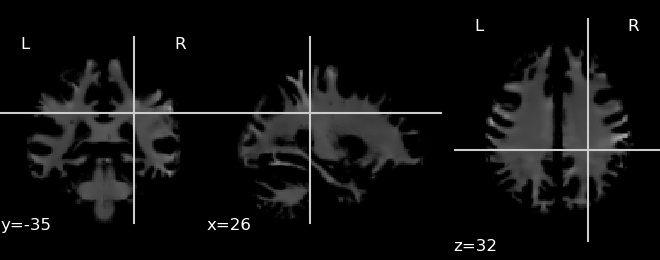}\\
        \multicolumn{1}{c}{\small (b) Extracted White Matter}\\
        \includegraphics[width=0.45\textwidth]{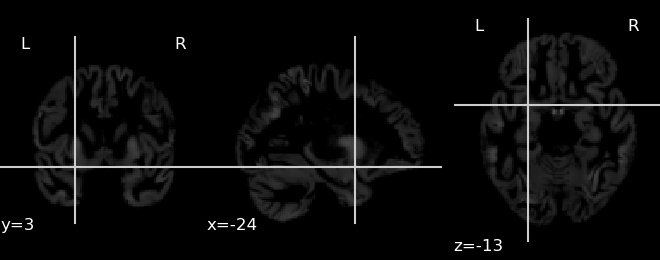}\\
        \multicolumn{1}{c}{\small (c) Extracted Gray Matter} {}   
    \end{tabular}
\caption{Sample MRI scans for a Healthy Control Patient and the extracted GM and WM}\label{fig2}

\end{figure}

The structure of our preprocessing pipeline is presented in Fig. ~\ref{fig1}. Since MRI intensity varies from subject to subject, to minimize discrepancies we normalize the values to a [0,1] range. After that the images were aligned to a standard space named Montreal Neurological Institute (MNI) and general intensity non-uniformities are removed using bias field correction (FAST)~\cite{biasfield}. FNIRT / BET~\cite{brainextract} was used to extract brain from the scans. The skull, fat and background regions which do not contain useful information were removed. The data was registered to MNI152 format (FLIRT)~\cite{registration,registration1}. After that artifact removal was performed, i.e. any voxel intensity values higher than 1 was corrected to be in the range [0,1]. Then a deformation method was applied to extract Gray Matter (GM) and White Matter (WM) from the scan and a 8mm Isotropic Gaussian Kernel was used to smooth and increase the signal-to-noise ratio and remove unnecessary portions of the scan. Finally we have three separate datasets: whole brain scans, GM and WM extracted from the brain and Smoothed GM and WM. Examples of the extracted brain and the resultant WM and GM extracted from the brain are given in Fig. ~\ref{fig2}. For our experiments, we did not use the whole brain scans, which produced less promising results based on our earlier analysis~\cite{firstpaper}. Our models were designed to handle the extracted WM and GM scans and the smoothed version of WM and GM scans.

\subsection{Model Architecture}
We created three separate ensemble architectures combining existing model architectures of the ILSVRC~\cite{imagenetcontest} implemented in Pytorch \cite{pytorch}. We selected 6 models to construct our ensemble architectures and choose the three architectures with better performances:
\begin{itemize}
  \item ResNet 101~\cite{resnet}
  \item SqueezeNet 1.1~\cite{squeezenet}
  \item DenseNet 201~\cite{densenet}
  \item VGG 19~\cite{vgg}
  \item MobileNet V2~\cite{mobilenet}
  \item ShuffleNet V2~\cite{shufflenet}
\end{itemize}
These models are available from Torchvision \cite{torchvision} in two versions: without any kind of training (untrained) and trained on the ImageNet dataset. We used both untrained and pretrained models to construct our ensemble networks and compared the performances of the resultant architectures to examine if training on the unrelated Imagenet dataset makes the models perform better in PD identification. Since the models were designed with the ImageNet dataset in mind, we had to modify the models in order to accommodate the format of MRI data. The input layers of all models were changed to accommodate the format of our input and the output layers were changed to predict between 2 classes (PD and HC) instead of the 1000 ImageNet classes. 
\subsection{Core Architecture}
\begin{figure}[htbp]
\centering
  \includegraphics[width=.5\textwidth]{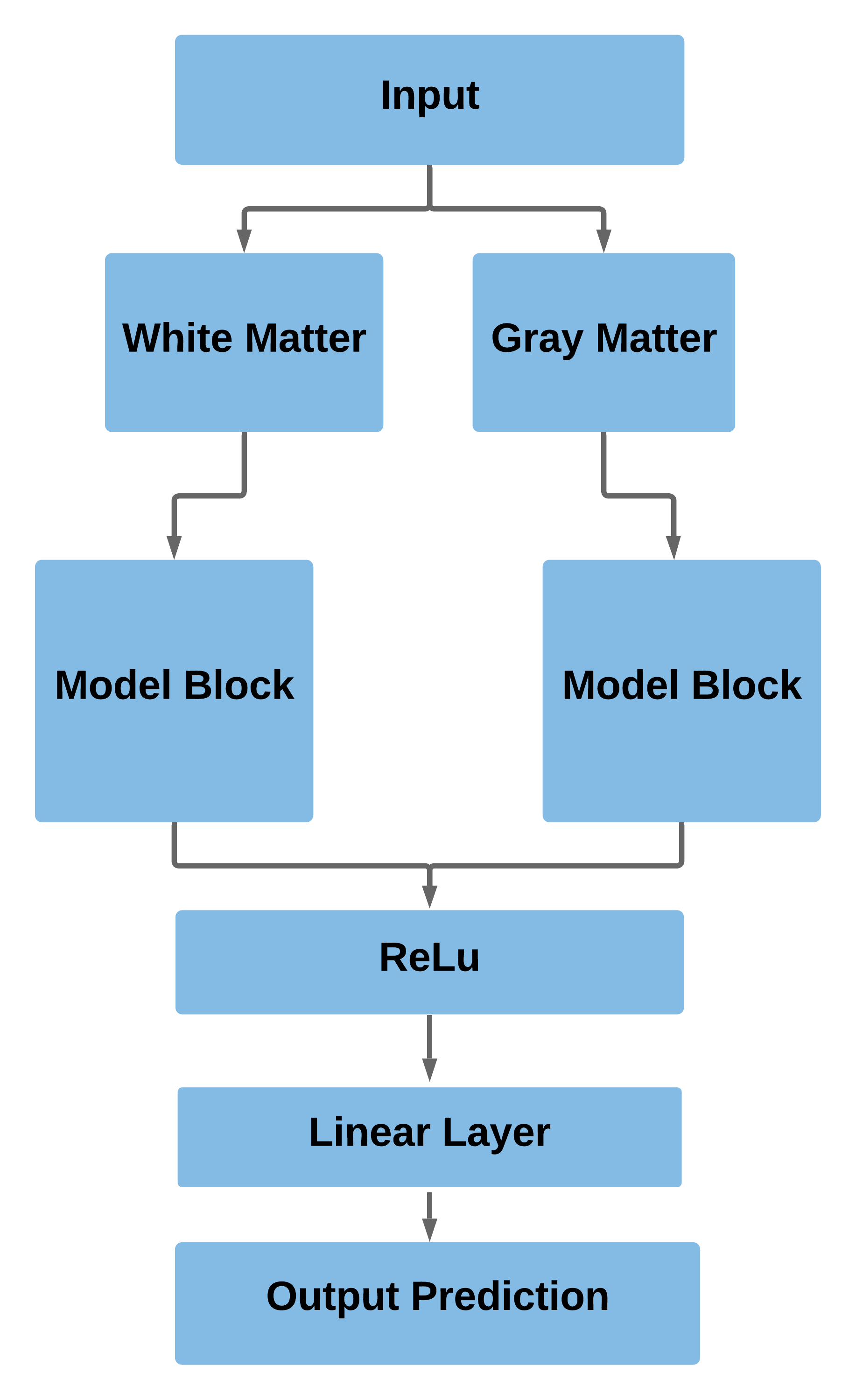}
  \medbreak
  \caption{Core Architecture}\label{core}
\end{figure}

For our main architecture, we take the extracted GM and WM scans dimension $121 \times 145 \times 121$ and pass them in parallel through two blocks comprised of multiple ILSVRC models as described above. We refer to this combination of models as a Model Block. We then concatenate the output from both blocks and then pass them through a ReLU activation layer and then through a final linear layer which predicts between the two output classes. 
Fig. ~\ref{core} shows a visual representation of this architecture.
We used multiple combinations of models to create different versions of Model Block, replaced them in the core architecture and tested their performances with both pretrained and untrained models.
\subsubsection{Architecture 1}
\begin{figure}[htbp]
\centering
  \includegraphics[width=.5\textwidth]{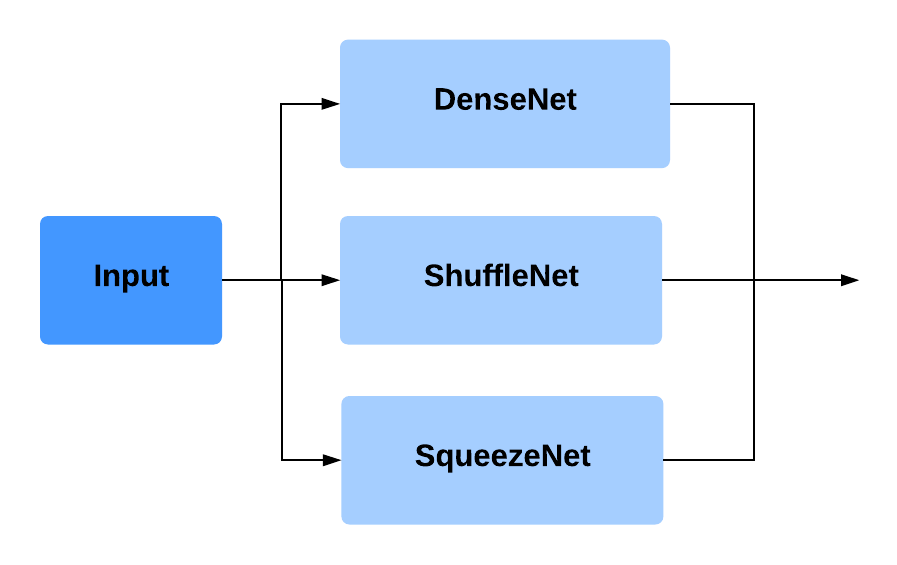}
  \medbreak
  \caption{Architecture 1 Model Block}\label{block1}
    \includegraphics[width=.5\textwidth]{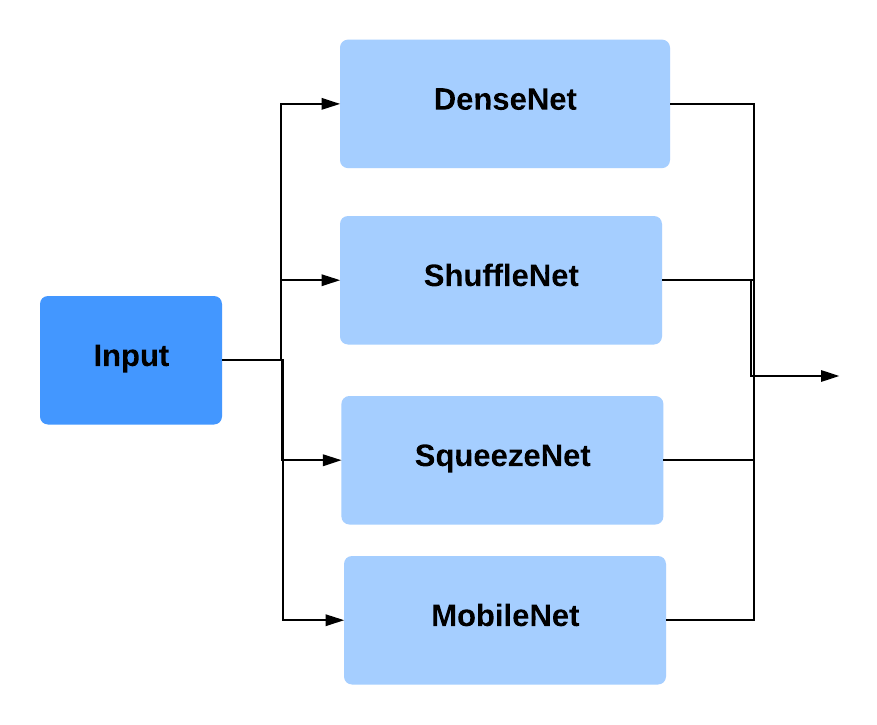}
  \medbreak
  \caption{Architecture 2 Model Block}\label{block2}
    \includegraphics[width=.5\textwidth]{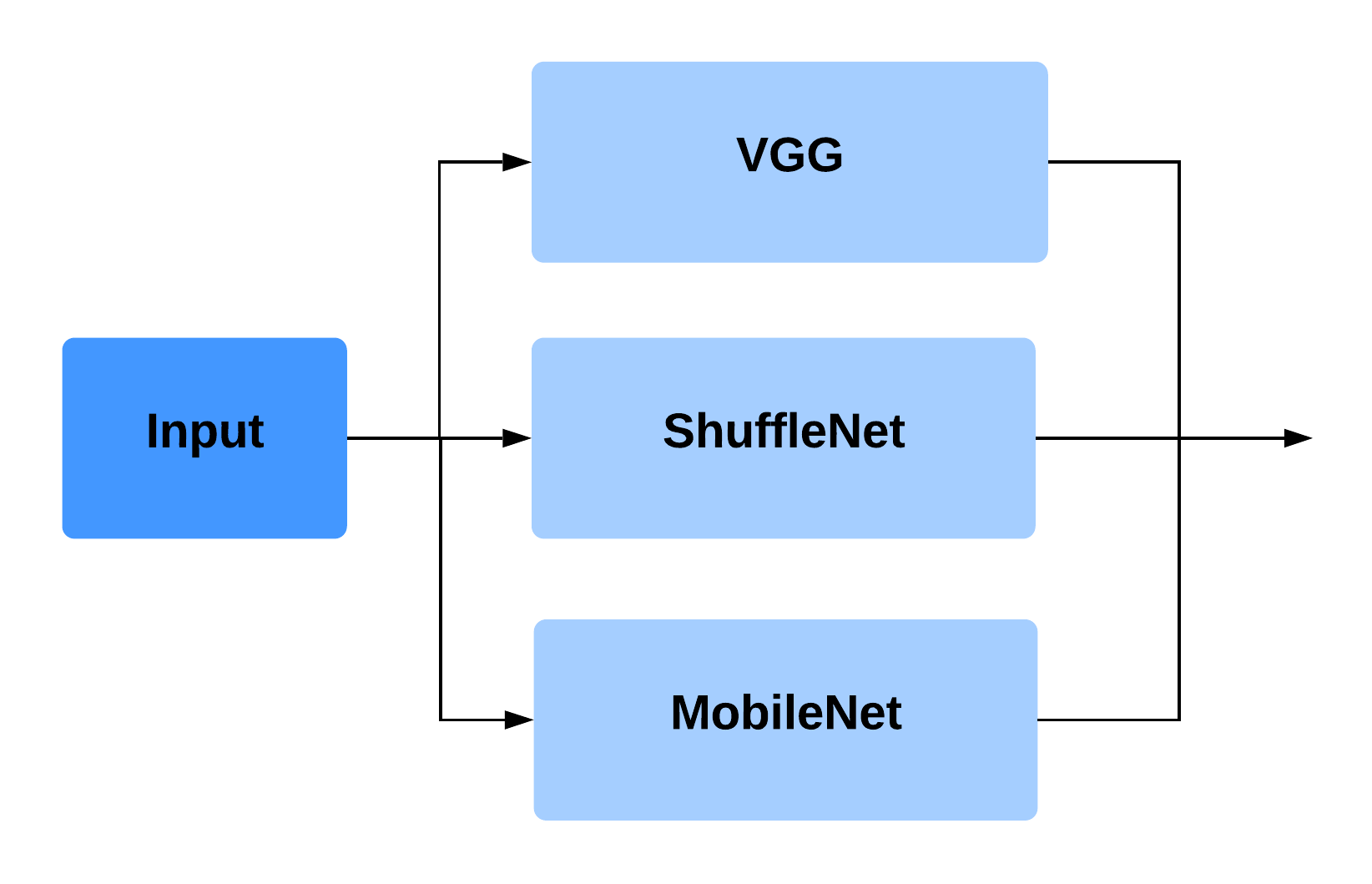}
  \medbreak
  \caption{Architecture 3 Model Block}\label{block3}
\end{figure}

The model block was comprised of DenseNet, ShuffleNet and SqueezeNet in parallel. The input was passed through all three models simultaneously, as shown in Fig.~\ref{block1}.

\subsubsection{Architecture 2}
The model block was created by adding MobileNet to Block 1, so it was comprised of DenseNet, ShuffleNet, SqueezeNet and MobileNet in parallel. The input was passed through all four models simultaneously, as shown in Fig.~\ref{block2}.

\subsubsection{Architecture 3}
The model block was created with ShuffleNet, VGG and MobileNet in parallel. The input was passed through all three models simultaneously, as shown in Fig.~\ref{block3}

\section{Experimental Results}
We created 2 separate versions for each of our ensemble architectures; one with all untrained constituent models and another with all pretrained constituent models. The dataset was randomly split and 80\% was selected for training and 20\% for testing. Each model was trained for 25 epochs with an Adam Optimizer. We used Cross Entropy Loss function. At each epoch, the training set was further split randomly and 20\% was selected for validation. We used a learning rate of \textbf{.001} and \textbf{.0001}. We repeated the process multiple times and the average results are presented in Table~\ref{resultstable} along with the results of some other approaches on similar data. 
\begin{table*}[htbp]
\centering
\setlength{\textwidth}{12pt}
\renewcommand{\arraystretch}{1.5}
\caption{Results}\label{resultstable}
\begin{tabular}{@{\extracolsep{\fill}} c | c | c | c | c }

 \thead{Model} & \thead{Use \\Smoothed \\ Scan} & \thead{Pre Trained} & \thead{Learning Rate} & \thead{Classification \\Accuracy\\(On a scale of 0-1)}\\
\hline
 \multirow{8}{*}{\makecell{Architecture 1}}&  \multirow{4}{*}{False} &\multirow{2}{*}{False} & .001 & $0.6045$\\\cline{4-5}

 &   & &.0001 & $0.8097$\\\cline{3-5}

 &   &\multirow{2}{*}{True} &.001 & $\textbf{0.9291}$\\\cline{4-5}
 &   & &.0001 & $\textbf{0.8955}$\\
  \cline{2-5} &  \multirow{4}{*}{True} &\multirow{2}{*}{False} & .001 & $0.7303$\\\cline{4-5}

 &   & &.0001 & $0.6592$\\\cline{3-5}

 &   &\multirow{2}{*}{True} &.001 & $\textbf{0.8315}$\\\cline{4-5}
 &   & &.0001 & $\textbf{0.7528}$\\\hline
  \multirow{8}{*}{\makecell{Architecture 2}}&  \multirow{4}{*}{False} &\multirow{2}{*}{False} & .001 & $0.5485$\\\cline{4-5}

 &   & &.0001 & $0.7276$\\\cline{3-5}

 &   &\multirow{2}{*}{True} &.001 & $\textbf{0.9515}$\\\cline{4-5}
 &   & &.0001 & $\textbf{0.9440}$\\
  \cline{2-5} &  \multirow{4}{*}{True} &\multirow{2}{*}{False} & .001 & $0.5768$\\\cline{4-5}

 &   & &.0001 & $0.7416$\\\cline{3-5}

 &   &\multirow{2}{*}{True} &.001 & $\textbf{0.8614}$\\\cline{4-5}
 &   & &.0001 & $\textbf{0.8352}$\\\hline
  \multirow{8}{*}{\makecell{Architecture 3}}&  \multirow{4}{*}{False} &\multirow{2}{*}{False} & .001 & $0.5187$\\\cline{4-5}

 &   & &.0001 & $0.5522$\\\cline{3-5}

 &   &\multirow{2}{*}{True} &.001 & $\textbf{0.9029}$\\\cline{4-5}
 &   & &.0001 & $\textbf{0.9254}$\\
  \cline{2-5} &  \multirow{4}{*}{True} &\multirow{2}{*}{False} & .001 & $0.5805$\\\cline{4-5}

 &   & &.0001 & $0.6105$\\\cline{3-5}

 &   &\multirow{2}{*}{True} &.001 & $\textbf{0.8914}$\\\cline{4-5}
 &   & &.0001 & $\textbf{0.8390}$\\\hline
 \multirow{8}{*}{\makecell{Our Previous Result ~\cite{firstpaper} \\ using extracted GM and WM \\Scans}}&  \multirow{4}{*}{False} &\multirow{2}{*}{False} & .001 & $0.5487 \pm 0.0002$\\\cline{4-5}

 &   & &.0001 & $0.6847 \pm 0.0093$\\\cline{3-5}

 &   &\multirow{2}{*}{True} &.001 & $\textbf{0.9231} \pm \textbf{0.0258}$\\\cline{4-5}
 &   & &.0001 & $\textbf{0.9366} \pm \textbf{0.0170}$\\
  \cline{2-5} &  \multirow{4}{*}{True} &\multirow{2}{*}{False} & .001 & $0.5410 \pm 0.0106$\\\cline{4-5}

 &   & &.0001 & $0.7276 \pm 0.0476$\\\cline{3-5}

 &   &\multirow{2}{*}{True} &.001 & $	\textbf{0.9291} \pm\textbf{ 0.0170}$\\\cline{4-5}
 &   & &.0001 & $\textbf{0.9470} \pm \textbf{0.0083}$\\\hline
 Focke et al.\cite{relatedwork1} [GM]&N/A&N/A&N/A&0.3953\\\hline
  Focke et al.\cite{relatedwork1} [WM]&N/A&N/A&N/A&0.4186\\\hline
   Babu et al.\cite{babu} [GM]&N/A&N/A&N/A&0.8721\\\hline
    Rana et al.\cite{rana} [GM \& WM]&N/A&N/A&N/A&0.8667\\\hline
        Rana et al.\cite{rana1}&N/A&N/A&N/A&0.8967\\\hline
\end{tabular}
\end{table*}

For all three different architectures, we can see that the accuracy increases significantly when we use pretrained models to construct the model blocks. For Architecture 1, the accuracy we achieved was $0.9291$ with non smoothed scans with a learning rate of $0.001$. For Architecture 2, we achieved a detection accuracy of $0.9515$ for non smooth scans with $0.001$ learning rate, which was the best overall accuracy we achieved in our experiments. Architecture 3 achieved an accuracy of $0.9254$ for non smooth scans and a learning rate of $0.0001$. All three of our architectures perform better than existing models on similar data and achieve above $90\%$ accuracy. For comparison, we also present our results from our earlier finding~\cite{firstpaper} where pretrained models provide higher detection accuracy as well. We note that we achieved better accuracy than our previous work with an architecture constructed with pretrained models and non smoothed scans. It is also noted that the smoothing process actually caused reduced detection accuracy in our experiments with these architectures.

\section{Conclusion and Future Works}
In this paper, we proposed three novel Ensemble architectures for Parkinson's Disease Detection, which outperform related works on similar dataset and achieve considerable detection accuracy of around 95.15\%. We also find that, when we used models pretrained on unrelated ImageNet dataset for the construction of the ensemble architectures, it significantly enhanced the performance on detecting PD compared to untrained models. Our finding suggests a promising direction, where unrelated training data can be considered when insufficient or no training data is available for a particular application.

Our next step will be to analyse the decision making process of our model, perform occlusion analysis to see which region of the scans our models use to make the decisions and use the Substantia Nigra as the target of significance to assess if it produces better results than using GM and WM regions.

\section*{Acknowledgment}
Special thanks to PPMI for supporting Parkinson’s disease research by maintaining and updating their clinical dataset.

Data used in the preparation of this article were obtained from the Parkinsons
Progression Markers Initiative (PPMI) database (www.ppmi-info.org/data). For
up-to-date information on the study, visit www.ppmi-info.org.

PPMI is a public-private partnership funded by the Michael J. Fox Foundation for Parkinsons Research and other funding partners listed at www.ppmi-info.org/fundingpartners.

Special thanks for the advice from Dr. Sara Soltaninejad (soltanin@ualberta.ca), previous PhD student at the Department of Computing Science, University of Alberta.

Financial support from the Natural Sciences and Engineering Research Council of Canada (NSERC) is gratefully acknowledged. 


\vspace{12pt}
\end{document}